# Introduction to discoverology

## *Piotr Homola*

## 1. Introduction

The road to technology leads through the application of basic science results, and basic science results are nothing other than ground-breaking basic science discoveries. Do you like your satnav system? Thank Einstein! Although humanity is seemingly doing extremely well at technological development, you can never claim that we could not do any better. Want to drive an anti-gravity vehicle as seen on your favorite Sci-Fi film? Sure, O.K., just wait for a corresponding discovery, apply the results and there we are. But wait... might this particular discovery have already been made? Is it possible that something that sounds so incredibly ground-breaking, like an anti-gravitation effect, could be missed by the community of science professionals? Yes, unfortunately it is. Simply because of how incredible it seems. Let us illustrate the point by reminding of one of the drawbacks in the process of acknowledging science results as discoveries. The drawback known as the Semmelweis effect[1] is rather poorly known among scientists, as it concerns a story that is shameful for the community claiming to be the elite of humanity. Dr. Ignaz Semmelweis (1818-1865) discovered that child-bed fever mortality rates reduced ten-fold when doctors washed their hands between patients and, most particularly, after an autopsy. He proposed washing hands between patients as a good practice in 1861, although he was unable to provide a scientific explanation. His hand-washing suggestions were rejected by doctors of his time, interestingly also for non-scientific reasons. For instance, some doctors refused to believe that gentlemen's hands could transmit disease. Semmelweis's discovery was widely accepted only in the early 1900's, nearly four decades after his death. Such a lack of acknowledgement of new knowledge is today known as the Semmelweis effect. A metaphor for a certain type of human behavior, the Semmelweis reflex-effect is characterized by rejection of a new knowledge because it contradicts entrenched norms, beliefs and paradigms[2]. After hearing the Semmelweis's story one immediately asks what can be done to avoid this and other similarly devastating obstacles to one's own independent thinking. One can imagine that, after hearing the first unpleasant reactions from his colleagues, Semmelweis gave up his attempts to disseminate his finding, simply to remain fully respected and to continue his career conventionally, without any revolutionary disturbances. How many more patients must then have died before another brave man discovered what we now consider the basic rule of hygiene?

---

[1]    M. Mortell, H. H. Hanan, E. B. Tannous, M. T. Jong, *Physician 'defiance' towards hand hygiene compliance: Is there a theory–practice–ethics gap?*, Journal of the Saudi Heart Association. 25 (3), 2013, pp. 203–208.

The Semmelweis example points to a non-scientific context of a discovery *post factum*. A more extensive discussion in this direction, based on an example of Einstein's and Poincare's discoveries, can be found in Grabinska[3], and to study the potentially existing anti-gravitation effect, the reader is advised to browse the name "Eugene Podkletnov" on the Internet. All these example considerations teach us that there exist mental obstacles to acceptance of ground-breaking discoveries once they are made.    One can imagine an extreme case where these obstacles turn out to be so strong that a major discovery is not recognized at all, preventing the human race from taking a completely new developmental path at a turning point in its history. A logical consideration of such an unfortunate situation makes us conclude that all possible discoveries are not necessarily made "sooner or later", it might matter *when* they are made, and sometimes they might be made *now or never*. Even by studying only popular culture one understands that, say, an asteroid doom threat motivates the fastest possible development along the *right* way, where *right* stands for the sequence of scientific activities leading to development of a technology capable of blowing up or diverting the asteroid. Spending the available scientific resources elsewhere, say, to research anti-gravitation, would, in this scenario, be a waste of time and energy deadly dangerous to the human race. In simple words: discover *the right things* first or perish... On the other hand, it is impossible to judge in advance which discovery or developmental path should be sought or followed first, which is the right plan. Nevertheless, a plan is a good thing, although the information we possess to inform a decision on which way to go is incomplete, we might want to optimize the use of our resources based on some wide agreement within society, including non-scientists, on the to-do list of what is important and desirable – to build a *human road map*. Here, we just conclude non-existence of such things as the road map that we, humans, sketch for ourselves. Despite existing visions, there is no core view upon which the human majority agrees, of what, how and when we are going to do with ourselves. To complete the logic of this article, let us propose *ad hoc* a human road map. In the shortest version, it should contain

1) the humanity survival "how to" and 2) the humanity well-being[4] "how to". It has to be stressed that the human road map considered here refers to humanity taken as a whole, and not to its subsets, like e.g. nations. We note, however, that a single human being should eagerly accept a global map as proposed here, understanding that an individual contribution to global human welfare is the condition *sin equa non* for individual success on a private road map, i.e. to the increase of individual quality of life. We also note that a religious approach to life is not excluded in such a mapping formulation: in this case the notion of the sum of welfare is considered in the context of eternity – personal God gave us the Universe "here and now", for some reason, and, since we were also given tools to improve in the given reality, let us better use them as well as we can. On the other hand, within non-religious thinking, one typically also considers the global *sum of welfare* a value worthy of individual effort. Thus, interestingly, human development could be, at least at first glance, considered a joint project of religious and non-religious humans, not excluding *a priori* anybody on the planet, thus enabling the largest possible manpower resources. Now let us consider point 1) on our human road map. As concluded from the asteroid example given above, we might want a diverse approach to our survival: a) developing technologies based on the *known* physics towards solutions preventing the human race from the demise caused by known cataclysms, and b) developing our understanding of the Universe within basic research in natural sciences by i) building theoretical models; ii) their experimental verification; iii ) practical implementation of conclusions; iv) staying ready and open to receive and acknowledge unexpected phenomena as hints to ground breaking discoveries, then, if such unexpected phenomena occur, coming back to point i).

In this article, we argue that humanity could do better, at least on point b). We have already argued that it is worthwhile to ask how to perform best on the way towards scientific discoveries, now we propose how to organize an effort in this direction. It seems that a methodology for making discoveries does not exist in a systematized, widely taught form, at least not to the knowledge of the author[5]. Although there exist interesting studies of human creativity and features of human psychology that lead scientists to discoveries, beginning with the classical study "Psychology of Invention in the Mathematical Field" by Jacques Hadamard[6], they are usually not aimed at systematically studying how to lead people to discoveries. Here, we propose such a systematization in the form of a discipline which would contains ideas, methods and activities oriented towards reducing logical, mental and organizational obstacles for novel thinking.

---

[4]    The term *humanity well-being* is used to name some yet not existing measure of global welfare and to stress its distinction from a similar expression: *quality of life*, a measure of happiness already very well grounded in sociology and often applicable to individuals and societies, although apparently not yet to the whole humanity.

[5]    The author is an astrophysicist with nearly 20 year of experience in the field. He has never been taught in a systematic way how to perform better in making discoveries. If such a methodology exists, it is apparently very poorly disseminated.

[6]    J. Hadamard, *An Essay on the Psychology of Invention in the Mathematical Field*. Princeton University Press, 1945.

How to *make* a discovery? One might say that to discover something you just need to be a genius or have some luck, nothing that would depend on the specific person who is to make the discovery. We consider this kind of thinking a myth and claim that a) one could be trained to increase the chances for making discoveries and b) the discovery training deserves the status of a discipline; we propose a distinct and meaningful name: *discoverology*.

The working question asked within this article concerns the obstacles one might encounter on the way to discoveries. If we are able to point to such obstacles and to the ways of avoiding them, it would give a net effect of improving the quality of scientific research, which would in turn speed up progress in understanding the Universe, and, consequently, the development of our civilization. If we can do any better at making discoveries, let us then do better.

## 2. Innovatics: the logistics of discovery

We note the literature practically addressing the environmental requirements supporting the process of thinking leading to innovations in applied sciences[7]. Optimum organizational behavior, multidisciplinary inspirations, information science and psychology of creativity can be listed as the key known components of a discovery(innovation)-friendly environment or toolbox. The latter component, psychology of creativity, provides, among other things not discussed here, some understanding of collaborative creativity. The bare fact of considering a collective and coordinated effort of many human brains in comparison to the performance of just one isolated individual suggest a potential for synergy in group thinking. Here we generalize the view of such potential in the perspective of challenges in basic science research: the more brains focused on solving a puzzle that Nature offers us, the better the chance of identifying a clue. Moreover, a proper formatting of the scientific issue to be considered might enlarge the circle of potential contributors, including non-professionals, and even children, as proved by a number of pro-jects under the "citizen science" flag[8]. The term "citizen science" is understood in a straightforward way: science must be real, i.e. yielding professional publications, meaning the participation of non-professionals (citizens) hand in hand with completely educated and experienced scientists is truly needed to improve the quality of scientific outcome or even essential to reach the science goals, as claimed e.g. by the Cosmic-Ray Extremely Distributed Observatory (CREDO) Collaboration[9]. While there are good examples of taking the citizen science research method as a valuable scientific tool (not outreach!) not only by single collaborations but also within well financed programs implemented by large consortia[10], the public is still not considered as an equal-right partner by the science community. Within the mainstream efforts on public engagement in science one most often finds proposals for a monologue, a one-way information flow (the public being *informed* well about what is going on in science) or, at best, a dialogue in which the public can deepen its level of scientific awareness[11], then get engaged on different levels in scientific processes, and even point to problems that require scientific treatment. The existence of well-developed and scientifically useful citizen science projects can be considered a proof of concept for engagement of the public as truly necessary scientific partners. This proof of concept offers prospects for a strategy towards an enlargement of the pool of talented young people enthusiastic about science and ready to complete their education in natural sciences. Thinking about civilizational development, one should, of course, propose a global citizen science strategy that would aim at optimizing the availability of human processing resources (brains and passion) for scientific research. Such a crowd-sourcing strategy should include:

a) Optimizing the social environment that would permit and stimulate personal development;
b) Providing fair information about opportunities and challenges in science and enabling attractive educational and career paths;
c) Offering a science taster by enabling participation by non-professionals;
d) Developing large scale social tools to support and make optimum use of collaborative creativity;
e) Introducing fun & motivation tools;
f) Providing discovery-oriented mind formation.

---

[7]     M. Jasieński, M. Rzeźnik, *Innovatics – a new toolbox of skills for innovative production managers* [in:] *Innovations in Management and Production Engineering*, Oficyna Wydawnicza Polskiego Towarzystwa Zarządzania Produkcją, Opole 2012, pp. 63-71., in book: *Innovations in Management and Production Engineering,* Chapter: *Innovatics – a new toolbox of skills for innovative production managers*, Oficyna Wydawnicza Polskiego Towarzystwa Zarządzania Produkcją, Opole, 2012, pp. 63-71.

---

[8]     http://zooniverse.org, as of 12.10.2017.
[9]     N. Dhital et al. (CREDO Collab.), *We are all the Cosmic-Ray Extremely Distributed Observatory*, PoS (ICRC2017) 1078, arXiv:1709.05196; http://credo.science, as of 12.10.2017.
[10]    ASTERICS: https://www.asterics2020.eu/, as of 12.10.2017.
[11]    National Congress of Science, Panel Upowszechnienie nauki i społeczna odpowiedzialność uczelni, Kraków, Poland, 2017, https://nkn.gov.pl/en, as of 12.10.2017.

Following this broadly defined discoverology road map, human civilization might expect that, on the base established by people enthusiastic about science, scientific (=developmental) capital will be built by cultivating the broadly seeded passion and letting it infect wider and wider circles and communities. Below, we briefly mention the proposals or already existing solutions going along with the global discoverology strategy:

## 2.1. Optimizing the social environment that would permit and stimulate personal development

A global effort to provide Internet access for everybody across the planet is underway, e.g. under the flag of the International Telecommunication Union (ITU)[12], with the phrasing "connect the world". Here we only note that, once the connection is provided, it has to be supplemented by some sense of usage. And the key sense and goal of a global connection as proposed within discoverology is to activate hitherto unused or under-explored human resources: brains. We want as many of them as possible to be dedicated and prepared for critical, independent and creative thinking. Thus, we propose that, in parallel to the technical efforts related to global Internet connectivity, also discoverological programs be developed, tested, and implemented prototypically in some medium-development regions – to get fully ready to and to benefit most from a globally available and affordable Internet connection, as soon as we have it.

## 2.2. Providing fair information about opportunities and challenges in science and enabling attractive educational and career paths

We propose that pro-scientific educational programs be oriented "top-down": beginning from an individual's fascination and discoverological enthusiasm about "top" scientific issues, i.e. the most challenging mysteries of the Universe in the macro and micro scale, through proposing short tutorials and mind formation towards a deeper understanding and involvement in the process of understanding these mysteries within the scientific methodology, potentially leading to a realization by participating individuals of the need to learn "downwards" the "boring & basic" details within traditional education, and finally ending with a new beginning: a tailor-made education profiled appropriately according to the initial, individual "wow". The gain is that, instead of asking talented individuals to become educated in the boring details that *might* be needed and only then telling them how to apply these details to something really exciting, "top-down" education is oriented on the details that *are* necessary to get involved into a specific, exciting scientific challenge. In such a model, the phase of excitement about science starts early – with the first scientific question in mind – and if the passion is properly cultivated, talented individuals might see the sense in undertaking the path of a scientific career. In the other model, with an overly elongated "boring" education phase, when one is kept away from investigations on the edge of the unknown until "fully prepared" by learning "all the basics", the talented young folk who are hungry for challenges *here and now* might escape to other fields of activity which are actually *less* challenging and demanding, but which offer opportunities earlier. The youngsters must then be well aware of the real opportunities and challenges in science *before* they take decisions determining their life paths. Top-down education which supplements traditional methods can then be one of the tools to enable optimum life-determining decisions, thus enlarging the pool of the most gifted members of the population in the most challenging fields of human activity which determine the development of the whole civilization. A good do-it-yourself beginning in a "top-down" education can be made e.g. by considering the compilation of the Universe's mysteries by Wikipedia: "List of unsolved problems in physics"[13]. A pure awareness of the existence of such a list is stimulating to an honestly thinking brain. A lot of fundamentally exciting challenges in science do exist! It is interesting to note that this simple reflection leads to a new complementarity or even a reversal of thinking in scientific outreach: we are prepared to receive the information stream dedicated to dissemination of knowledge, while it might make sense also to disseminate *awareness of missing knowledge* – awareness of exciting scientific challenges waiting for the most talented human individuals.

---

[12]    http://www.itu.int, as of 12.10.2017.

## 2.3. Offering science taster by enabling participation by non-professionals

One refers to the already mentioned citizen science projects as opportunities for valuable participation in scientific research after just a very short on-line tutorial, compatible with the "top-down" education model. Notwithstanding that quite a few citizen science projects exist and are developing very well, only one of these seems to offer an easy and global-citizen-appropriate science format for both data acquisition and analysis. This is the already mentioned CREDO project: a newly launched worldwide collaboration aiming at global research on data available everywhere and for free – cosmic rays. The research proposed by CREDO addresses fundamentally important science issues like interrogating the fine structure of space-time, verifying models for Dark Matter or observing possible manifestations

---

[13] https://en.wikipedia.org/wiki/List_of_unsolved_problems_in_physics, as of 12.10.2017.

of quantum gravity. While deep and ambitious scientific objectives are the core motivation for the CREDO Collaboration, the non-scientific potential, e.g. in the area of outreach and education, cannot be considered a mere side activity. Rather, it is rather a basic tool to attain the main scientific objectives, which seems to be a unique feature for a scientific project. Let us explain in some more detail the importance of public engagement in CREDO. An easy method being implemented in the project to engage the public in data acquisition is through smartphones: popular devices which can serve as detectors if equipped with a special application[14]. The next level of engagement proposed by CREDO is not much different from other citizen science projects: a web-based interface named Dark Universe Welcome[15] that enables an easy visual and crowd-sourced classification of detection patterns in a globally-spread network of detectors. The network, in order to yield valuable scientific results, has to involve as many active cosmic detectors and experiments as possible, including the largest professional cosmic-ray arrays, gamma-ray telescopes, satellite detectors, neutrino observatories, accelerator detectors (no beam mode), educational devices and the most popular scientific devices of all: smartphones. As explained in the definition of the CREDO mission, there are two main research channels: a) verification of existing theoretical models based on which cosmic ray ensembles (CRE) are expected on Earth; and b) the quest for unexpected physics, beyond the current models and perhaps even the imagination of theorists. Each of the tests performed in channel a) will impose specific detector requirements to optimize the scientific quality. Once such requirements are known, a decision can be taken on whether one could use only a subset of the existing CREDO stations or whether additional construction is needed. In such a situation, one might ask how many stations of which type and of which geographical spread are required to give sensibly valuable results, either in terms of observation or to apply upper limits. On the other hand, a research located in channel b) needs "only" the biggest possible scale, in terms of the total collecting surface of the whole network, spatial distribution of its components and the number of participants ready to analyze the data. With this remark, one realizes that the bigger CREDO is, the higher the chances for discoveries in channel b). Therefore, public engagement, together with standard outreach, can be considered basic tools to implement the CREDO scientific strategy. Consequently, one understands that even a minimum action that leads to collection of just one more cosmic ray particle deserves a share in sci-

---

    *Piotr Homola*

entific publications using the CREDO data and, potentially, also in the scientific awards. One can easily imagine that a single click on the CREDO Collaboration fan page might attract new participants who could either provide more data or help analyze the already acquired resources, or both. Then they can get more and more deeply engaged by entering the available paths of self-development connected to the project, including higher education in natural sciences. Then, effectively, each symbolic "like" should give membership and co-authorship rights in the CREDO publications!

## 2.4. Developing large scale social tools to support and make optimum use of collaborative creativity

Although existing social media solutions already provide a very good starting point for a global creativity effort, they do not seem to be capable of driving a fully functional multi-million budget scientific environment. We suggest that a list of desired functionalities is evolving naturally along with the development of the community, and the starting items might include a smartphone detector-analysis interface, project-friendly solutions like a worldwide, user-fueled system to evaluate scientific competences and related skills, to facilitate browsing for appropriate project members to complete optimum teams. In principle, such a scientific environment-framework can be distributed worldwide, generating synergies by using the resources provided by participating institutions and business partners. A prototype of such a widely distributed system supporting (mostly young) enthusiasts of research in natural sciences is already being developed as the Incubator of Scientific Discoveries[16] (see Fig. 1 for the schematic structure).

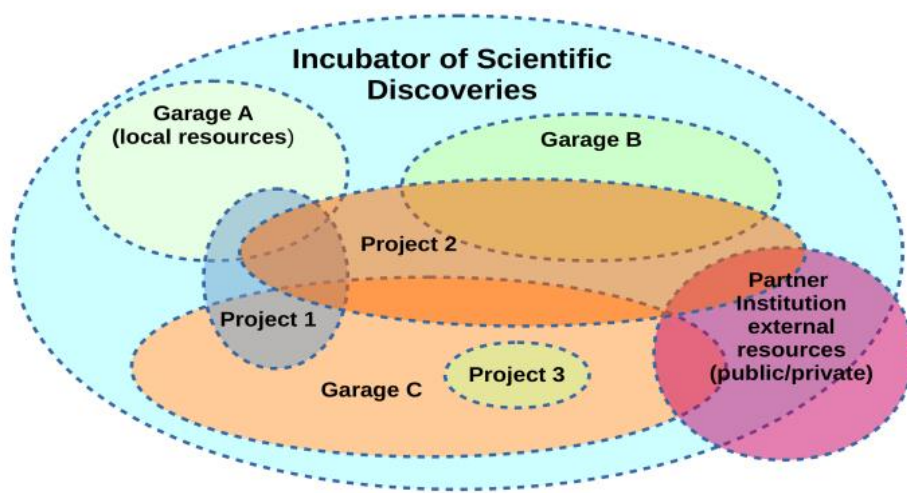

*Fig. 1. Schematic structure of the Incubator of Scientific Discoveries: discoverology-friendly environment dedicated to young science enthusiasts.*

---

[16] http://incubatorof.science, as of 12.10.2017.

## 2.5. Introducing fun & motivation tools

Considering the popularity of smartphone-related activities like Pokémon games or Treasure Hunts, one might expect the CREDO concept to possess the potential to activate similar mechanisms, as a source not only of fun but also of useful work for top science objectives. Instead of hunting for items assigned to stable locations within the CREDO detection framework, one would invent strategies to build a sufficiently populated and appropriately configured group of smartphone users (see e.g. CRAYFIS[17], DECO[18], or CREDO Detector[19]) or operators or some other popular detectors (see e.g. Cosmic Pi[20] or Cosmicwatch[21]) to hunt for extensive air showers. The larger the group, the larger the likelihood of catching more prominent events: rare showers induced by particles of the highest known energies. The more prominent an event the group would detects, the more ranking points would be awarded. One naturally understands that the higher the individual or group ranking, the higher the share in the scientific results. In the case of the CREDO mission, apart from detecting known phenomena like extensive air showers, one would be interested even more in large scale temporal correlations of such events. Such an interest should stimulate the exchange of thoughts and experiences between distant groups, both in the geographical and cultural sense. Moreover, the seismic, geomagnetic and cosmic weather monitoring that is to be implemented within the CREDO network would add an important social dimension to motivate the user community: hunting for multi-channel coincidences might provide a way to predict major earthquakes, thus saving human lives! Although the existence of such coincidences is hypothetical, a discussion in the professional literature[22] provides motivation for economically feasible tests of this hypothesis. The main reasoning concerns globally visible transient changes in the geomagnetic field on and above the surface of the planet that could be induced by mechanical movements of the liquid iron in the interior of the Earth, keeping in mind that this internal liquid iron is assumed to be responsible for geomagnetism. The massive transient movements in the interior of the Earth could potentially initiate seismic effects that would propagate from the Earth's outer core to the crust at lower speeds such that the accompanying geomagnetism changes, and these geomagnetic changes could induce variations of the low energy cosmic ray rates detectable on the surface of the planet. We note that the CREDO network will

---

be the largest and the most efficient instrument to verify such a hypothesis, and even if the chance of finding a method of predicting earthquakes with sufficiently good precision is very small, it will be a serious responsibility for the CREDO community to make an effort towards either confirmation or exclusion of the hypothesis. Once various rankings become available for CREDO users, the fun can be maintained by organizing events like art-science happenings, conventions, detection tourism to sites where detection probability is higher, e.g. nearby large, professional observatories connected to the global network which provide an additional trigger confidence, i.e. everything that gives more ranking points.

## 2.6. Providing discovery-oriented mind formation

Once one realizes that the pool of talented colleagues passionate about science is limited only by the size of the human population, it is time to consider an optimum way of using their collective thinking potential – to reach the highest possible synergy gain. It means intellectual formation in a form which cannot be known in advance. What is clear is that, once the community of science enthusiasts is active and ready to develop, it must be given the relevant opportunities. Here we enter the landscape of *discoverology*, which is not a totally blank canvas, although it requires serious painting effort. Since *discoverology* does not seem to exist as a discipline, and since at least some trial efforts to facilitate scientific discoveries can be imagined, we propose to begin with this initial imagination and try to reach the grown-up phase of the subject with a method of successive approximations. Thus, we initiate the training by realizing the need for improvement in discoverological thinking. Then we try to identify obstacles and propose ways of overcoming them, constantly keeping in mind the necessity of exchanging experience and conclusions with the other trainees. The first approximation of *discoverology* as a discipline has already been made. The reader is referred to a series of mini lectures on this subject given by the author of this article at meetings of the Incubator of Scientific Discoveries[23]. Below, a short summary of the outcome of these lectures is summarized by describing the three possible sub-disciplines of *discoverology*: *choiceology*, *errology* and *questiology* (see Fig. 2 for a diagram).

# 3. Choiceology

*Choiceology* is partly fueled by *innovatics* as explained above: once the scope of opportunities is available, one has to make a choice of direction, and we would like that this choice be based purely on intellectual considerations. Here let us consider a fundamental choice: should we, humans, perform scientific research at all?

---

[23] http://credo.science/homola/discovery-thinking, as of 12.10.2017.

Somehow we do, probably unintentionally, and probably just because we are capable of doing science. Once upon a time, we tried, and it proved to bring us benefits and welfare. Although we were most likely not aware of the situation of choice at the beginnings of science, we should by now be aware that we have the choice either to a) continue scientific research or b) to stop it. The answer seems to be obvious: a). By considering the philosophical context here one might point to the distinctly different ways of reaching the answer: again we propose a distinction between a religious non-religious approach. In the former case, once one assumes the existence of the Creator, the logical consequence is to also take into consideration the plan the Creator might have for the environment He created and for its inhabitants, and also the possibility that the Creator provides tools and messages of different form but sufficient to follow the big plan which, by the way, should be a best scenario for the community of inhabitants as a whole. Then the Universe should be considered the Book, having the capability of efficient scientific investigations – a tool to read the Book, and the fact of the existence of the Book and of the reading tools implies a necessity to read, i.e. to do scientific research and take the outcome seriously. On the other hand, within non-religious thinking one should also be dedicated to basic science, at least through consideration of the practical welfare of the whole community which is then transposed to an increase in the major non-religious value: individual welfare. Thus, basic scientific research seems to be a universal, unifying value, something worth cultivating in a well-planned way, and this well-planned way can be nothing else but the *discoverology* road map discussed here.

Interestingly, after making the choice of not giving up scientific activity and looking around with our discoverological road map, we realize the existence of bigger or smaller crossroads which are not indicated on the map and not even discussed with the available methodological toolbox: these are our everyday scientific choices – the subjects of *choicological* considerations. One of the basic questions each scientist should ask him/herself is about the relation of the "volumes" of *known* and *unknown* in natural sciences: $R := V_{known} / V_{unknown}$. We note that the assumption about R determines scientific paths to be undertaken by an individual scientist. Assuming $R \gg 1$ we have to believe the standard models of physics and focus on how to reconcile the apparent theoretical and experimental inconsistencies to build a complete picture of our reality with the known blocks. On the other hand, if one admits the possibility that $R \ll 1$, this would imply thinking beyond the scientific paradigm and taking the known inconsistencies not as stoppers, but rather as potential boosters of scientific development: like windows to the unknown. Then we also tend to assume the existence of some hypothetical blocks of the big Nature puzzle, the blocks that help to complete the overall picture. In consequence, we undertake investigations which are focused to study firstly these "more promising", according to our *feeling*, windows to New Physics, verifying empirically the existence of "extra" hypothetical blocks. This differs from the "$R \gg 1$" approach, where one rather tends to treat the apparent theoretical and experimental inconsistencies as effects of secondary importance or artifacts.

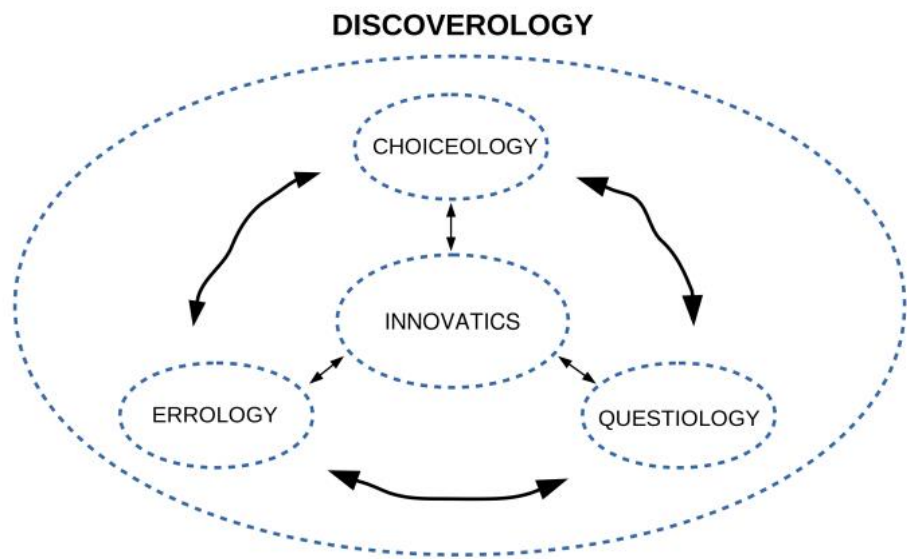

*Fig. 2. Logical structure of discoverology.*

In any case, it is clear that an assumption on R, which is apparently a free decision of a human mind, determines the scientific focus of an individual researcher and hence affects in some way the efficiency measure of scientific efforts taken as a whole: the area of the *land of the known*. The choice, although intellectual, does not seem to be grounded in scientific methodology: it is based on very limited information – by definition the volume of *unknown* cannot be determined with the toolbox composed of *known* methods. Thus, a scientifically very important choice between $R \gg 1$ or $R \ll 1$ (for clarity we do not discuss here the whole possible spectrum of R) is beyond the competencies of scientists. Here the *discoverological* approach seems to be in place: can one do something to increase the discovery success rate, i.e. to enlarge the amount of information useful in the decision-making process? Let us consider, for instance, a discussion presented elsewhere[24], showing that religious thinking might introduce a tendency to think and investigate more towards $R \ll 1$ scenarios. While this does not mean that a scientist must consult a theologian before formulating the research plan, it might indicate some yet unexplored areas of philosophical inspiration for science. One such inspiration taken from the non-religious side can concern economical aspects of scientific discoveries. Let us illustrate this example with Fig. 3. Consider a scientific goal defined as observing physics beyond the Standard Model of particles (Standard Model) in cosmology. Such a goal is often referred to as "New Physics". Such an observation is commonly expected, it should give us the chance for a deeper understanding of the Universe, and hopefully a breakthrough leading to further development of our civilization. Now consider various ideas and projects that are scientifically very well motivated, i.e. possibly and hopefully leading to New Physics. Some of the proposals are cheap, some expensive and some extremely expensive; in addition, some are more complicated than others and some make different assumptions about R than others. Which way should we proceed? The majority of scientists would probably propose: let us try each of the ways, if resources permit. But at the same time, we scientists are all aware that the available resources, both in terms of manpower and funding, are utterly insufficient to reflect our scientific imagination and appetites. One simply does not receive grants whenever one applies. So what is the path to success? Which additional information, external, with respect to the available scientific methodology, can help make a scientific choice which would be optimal from the point of view of society as a whole? Since the question address-es the whole society, to give an answer one obviously needs to look at society as a whole and take into account the global needs and priorities. One might ask, for in-stance, which of the projects humanity needs more: an "interplanetary accelerator" or a "solution permanently eliminating child deaths from hunger"? What if some of the children who do not die of hunger thanks to not funding the interplanetary accelerator would instead become ground-breaking discoverers, and their discoveries after all help us learn about New Physics without the need to build a new accelerator? Nobody can answer this. Maybe learning about New Physics *faster* would increase our chances to survive some apocalyptic catastrophe, e.g. by teleporting through a wormhole to a better universe. Consider another question: if we have to implement those projects aiming to drive New Physics one by one, not all in parallel, what should be the sequence? Shouldn't we start with the least expensive and the easiest project? We must be aware that such considerations have no scientific background and, since the choice must somehow be made, the community of scientists must ask "externals" for help. With this article, we postulate that these "externals", i.e. philosophers, be consulted more often by investigators focused on basic sciences, also in fields like astroparticle physics: such consultation also provides ready opportunities to ask other types of questions, including ethical ones. This could help to practically evaluate the usefulness of *choiceology* – we could see "by experiment" whether adding some practical philosophy to our standard scientific toolbox makes a visible difference.

---

[24]    P. Homola, *Experimental way to the Theory of Everything, or can a theologian inspire a physicist?*, [in:] Częstochowski Kalendarz Astronomiczny 2017, Astronomia Nova & Wydawnictwo Akademii im. Jana Długosza w Częstochowie, 2016, pp. 213-224.

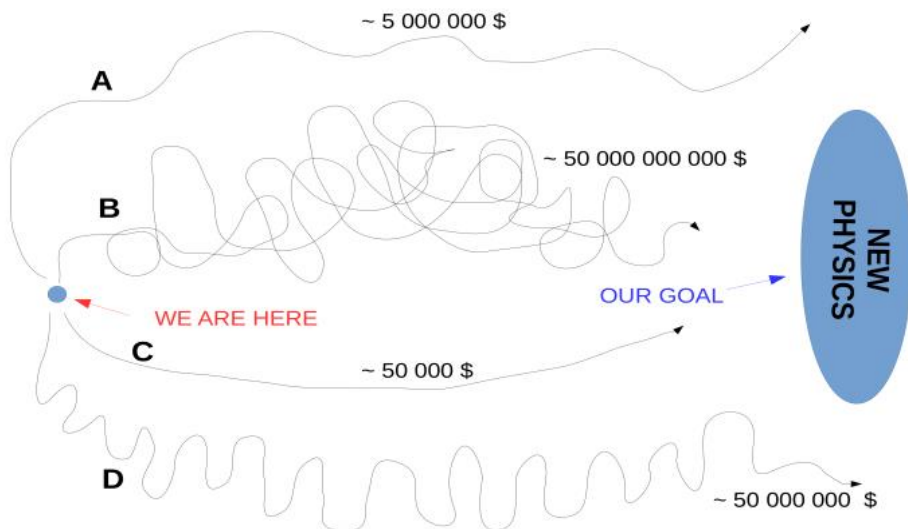

*Fig. 3. A choiceological example of economy of a discovery.*

## 4. Questiology

*Questiology* should teach us that unasked or abandoned questions might mean missed discoveries. Some questions might remain unasked just because the individual's thinking perspective is too narrow, under the overwhelming rigor of the available scientific paradigm, and some might get abandoned just because they are considered "stupid", either "internally", after some self-reflection, or "externally", e.g. by some well-established authority. Apparently, one can plan psychological training on how to be brave enough to ask a question "externally" – to a colleague or to a wider public, getting exposed to criticism and risking being classified as stupid. This risk, however, also brings opportunity. A public expression is a weight probe of the quality and potential of the question being asked, and the received feedback, whatever its form, if considered with care, increases the chances of a proper decision at the moment of choice: continue or abandon. So individuals who tend to underestimate their intellectual capabilities should express their questions in public, because this might be the only way to gain sufficient encouragement to continue the investigations they have in mind. On the other hand, those who frequently overestimate the quality of their own thinking should benefit from expressing their ideas in public and considering the received criticism, because this could help to save their time: they would continue their research only in those directions which pass the public quality test. While we agree that public verification of one's own ideas makes sense, we must be aware of possible obstacles that could mislead us and discourage our brave and novel thinking. One such obstacle is superior quality in the paradigm with respect to which the novelty is going to be proposed. If the paradigm is too good, it can become opaque, as noted by Kuhn[25]. Being blocked by an opaque paradigm is to suffer from a physical inability to accept any beyond-paradigm ideas as sensible. – Another issue might arise if one addresses a question to a community dominated by a super-strong, super-experienced and super-clever authority, whose "intuition" indicates that the proposed idea does not make sense. As long as this is only "intuition", i.e. a non-scientific tool for making scientific choices and judgments, one has to be very careful by accepting such an "opinion" as a definitive judgment. The obstacle of an overwhelming authority is illustrated with Fig. 4.

---

[25] T. S. Kuhn, *The Structure of Scientific Revolutions,* (2nd Edition) University of Chicago Press 1970, p. 85.

Among other questiology examples one notes also purely logical considerations based on an elementary hygiene and honesty of thinking: conclusions drawn based on certain assumptions can be questioned if any of these assumptions can be questioned, and any assumptions are questionable by definition. So it should be a good *questiological* practice to ask from time to time "what happens if I alter one of the assumptions?". Such a need particularly concerns conclusions that become foundations for whole branches of research programs. One good example is the mystery of Dark Matter. A tremendous worldwide scientific effort is ongoing to understand what Dark Matter *is*, we all want to observe it somehow. On the other hand, the *questiological* approach would push forward the question of whether Dark Matter exists at all. Luckily, this question has already been asked in public by proposing a theory of emergent gravity which has the potential to explain all the observational evidence for "Dark Matter" without the need to introduce any mysterious new component of the Universe[26]. Another *questiological* example concerns the key motivation of the CREDO project described above. The paradigm of ultra-high energy cosmic rays includes strong statements about non-observation of ultra-high energy photons[27] which can be transposed into the "fact" of the non-existence of ultra-high energy photons. The non-observation conclusion is based, however, on a set of fundamental assumptions coming from extrapolation by many orders of magnitude of the physics laws known from accelerator experiments. One such assumption is that the effect known in the literature as photon decay[28], where lifetime of an ultra-high energy photon might be as short as 1 second, does not occur in reality. The photon decay effect is a specific manifestation of violation of a very important and standard principle of physics: invariance of Lorentz transforms. Violation of Lorentz invariance is expected in models concerning Grand Unified Theories, and in the quest for a deeper understanding of the laws ruling our Universe one often assumes that a theory that unifies the known interactions exists. On the other hand, most of the models describing Lorentz invariance violation pre-

                        *Piotr Homola*

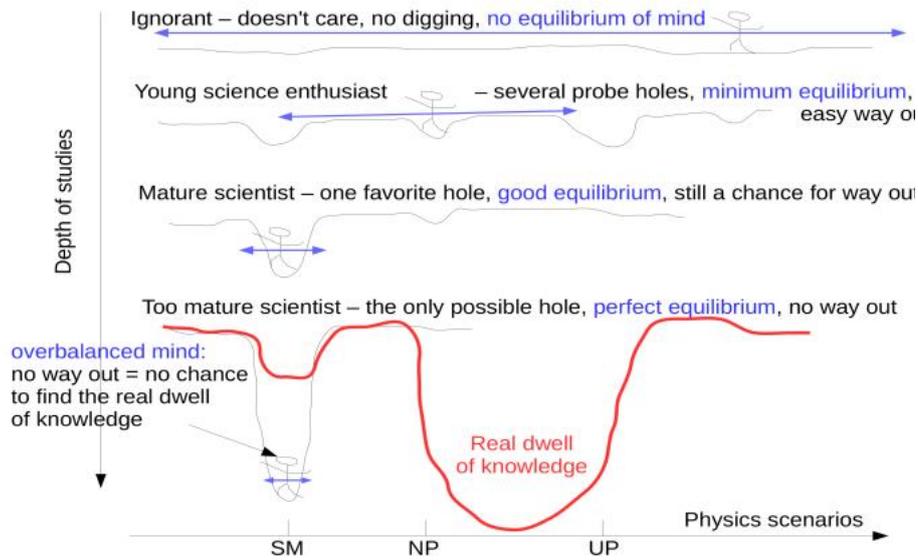

*Fig. 4. Dicoverological obstacle of an overwhelming authority.*

dicts the increase of ultra-high energy photon lifetime or mean free path. Therefore, the assumption about non-occurrence of the photon decay effect is reasonable, it belongs to the paradigm thinking, although it is not theoretically excluded. Let us now consider the following question: what if the photon decay effect *does* occur in reality? We note that, in this case, ultra-high energy photons have no chance to reach Earth, which is in trivial agreement with the observational limits. Consequently, and even more methodologically meaningful, if ultra-high energy photons have no chance to reach Earth, non-observation of ultra-high energy photons on Earth *cannot* be interpreted as non-existence of ultra-high energy photons. This is a trivial note which, however, carries serious consequences. Within the paradigm thinking, one does not consider ultra-high energy photons to be possibly existing physical objects, and this results in stopping the development of the models where ultra-high energy photons could be produced, e.g. the so-called exotic (or top-down) models[29]. On the other hand, admitting the photon decay effect (or other mechanism preventing ultra-high energy photons from reaching Earth) into consideration leads to a practical experimental conclusion: if ultra-high energy photons cannot be observed because they quickly decay into large cascades of secondary electromagnetic particles, then the only method by which to confirm their hypothesized existence would be to look for cascades of cosmic rays – complementarily to the state-of-the-art cosmic-ray research focused on detection and studies on single, uncorrelated particles (see Fig. 5 for an illustration). If, in the course of cosmic-ray research, we begin to think also about detection of ensembles of correlated cosmic particles, it becomes clear that we are entering a scientifically untouched territo-

---

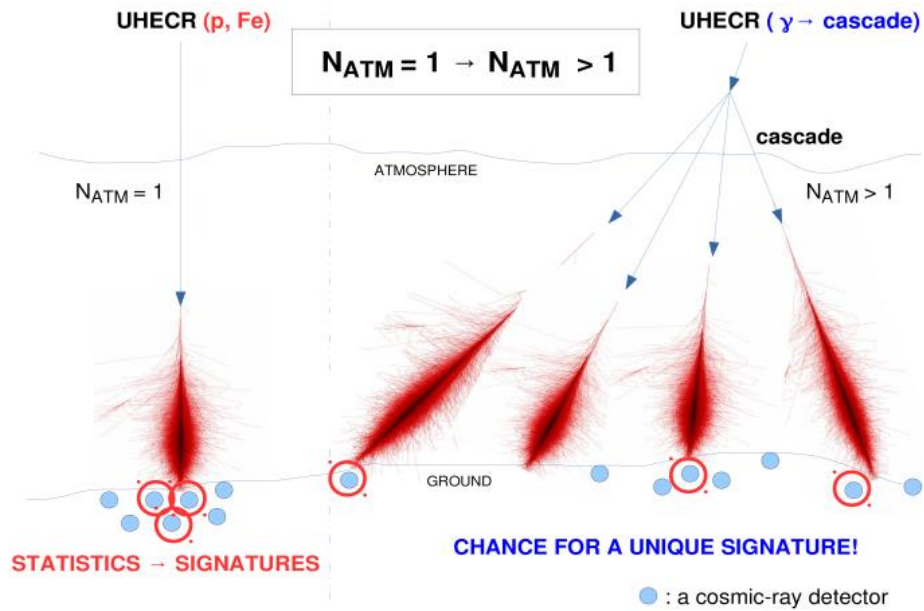

*Fig. 5. A generalization of the state-of-the-art cosmic-ray research by extending the investigations over cosmic ray ensembles.*

ry – opening a new window to the Universe, the window of cosmic ray ensembles (CRE)[30]. Thus, one may conclude that *questiology* proves its worth: we obtain an example where the questioning of one of the paradigmatic assumptions in the field of astroparticle physics takes us to a scientific *terra incognita*, giving new breath to a large class of theoretical models, as illustrated in a schematic form in Fig. 6.

## 5. Errology

*Errology* can probably be seen as the least controversial sub-discipline of *discoverology*. We might expect that each complex work performed by a human being might contain internal errors which are difficult to discover at first sight, and some might even be hidden forever. This holds for scientific achievements as well. When drawing conclusions of importance for civilizational development, it would therefore be good at least to estimate the risk that might potentially result from simple mistakes in our basic assumptions or tools. By simple mistakes, we understand the operations which we know how to perform correctly but somehow, fail to do so. Let us give a very simple example here: 2 + 2 = 40. We know the outcome of the operation should be 4, but, due to some absentmindedness of any origin, we add zero at the end, quite unconsciously. We are so sure about the 4 at the end that we might reduce our concentration which we typically apply for math operations, and write the result pretty automatically, without checking it. Now imagine that our example operation is a part of the process leading to a serious and unique result, something that happens frequently in novel scientific research on the edge of unknown, i.e. without any standard benchmark to compare with. Then, of course, if this unique and important result is taken uncritically by the community of experts, or if it cannot be repeated for any reasons, for instance economical ones, than the internal errors hidden in our work will propagate with the help of colleagues who will apply our results in their subsequent investigations. We want to avoid such propagation, and *errology* should define a program to systematize an anti-error effort within the scientific community. Such an effort should lead to the setting of basic standards for error-checking, leading to development of some kind of certificates assigned to each scientific work – something like the energy consumption ratings assigned to washing machines and other household appliances. In any case, we should not assume *a priori* that all published scientific results are correct. Of course, it is hard to imagine doing all studies at least twice by independent teams, although one can imagine repeating the *essential* results, where the risk of wide propagation of the possibly existing mistakes is large and might have serious consequences. An *errologist* would be an expert trained to identify such essential works and to point to their weak points.

---

30    P. Homola et al. (CREDO Collab.), *Search for Extensive Photon Cascades with the Cosmic-Ray Extremely Distributed Observatory*, in CERN Proceedings, PHOTON 2017 Conference (submitted).

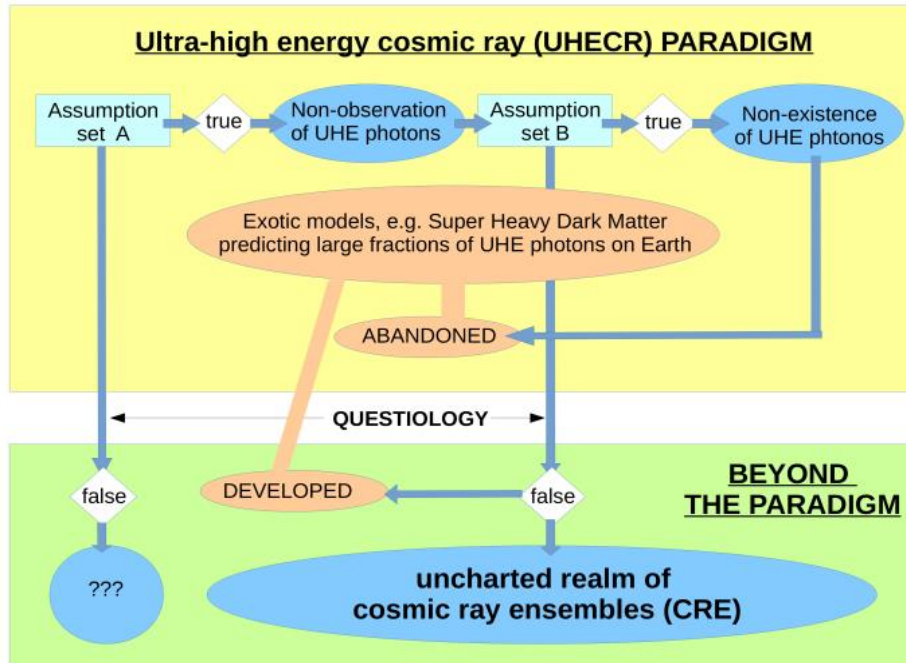

*Fig. 6. The thinking scheme leading to the formulation of the novel research program oriented on cosmic ray ensembles.*

Even if we can't afford to double-check all the potentially weak points of our scientific basis, the ability to estimate the risk resulting from "stupid" mistakes could motivate easy checks, even self-checking, that would already bring an additional quality to the scientific efficiency considered globally. So far, we can count only on scientists' self-responsibility, which cannot be quantified in any way. This lack of quantification introduces an additional, although avoidable uncertainty in the scientific results. As a result, our overall scientific progress is undoubtedly slowed down and some developmental stages might be even blocked permanently.

A few good *errological* examples come from the experience of the author. We list here a "factor 2" mistake in the standard reference used by researchers studying the behavior of ultra-high energy photons above the Earth atmosphere[31] (see Fig. 7 for an illustration), discovered by repeating the calculations presented in the paper[32]. Another example is the unexplained "factor 2" discrepancy in the simulated size of an electromagnetic cascade initiated by an ultra-high energy photon above the atmosphere seen by comparing outputs of the only two public and standard programs capable of performing such a simulation[33]. The issue remains unresolved because of the lack of interest of the developers of one of the programs. We also mention five "stupid" mistakes found in the code prepared for one of the largest cosmic-ray experiments to simulate azimuthal anisotropy in the distribution of Cherenkov photons in extensive air showers: "plus instead of minus", "floating point too much left/right", etc. These mistakes were found because the code to perform the simulations was written independently by two people: the prototype was coded by the author of this article and the final version by a skillful programmer, also a scientist, but capable of meeting coding standards. Since it was expected that both versions would give exactly the same output, the noted discrepancies were carefully tracked back to identify the issue. Interestingly, the mistakes were found in both versions, two in the prototype and three in the standard-meeting code, which might suggest a universality of the problem.

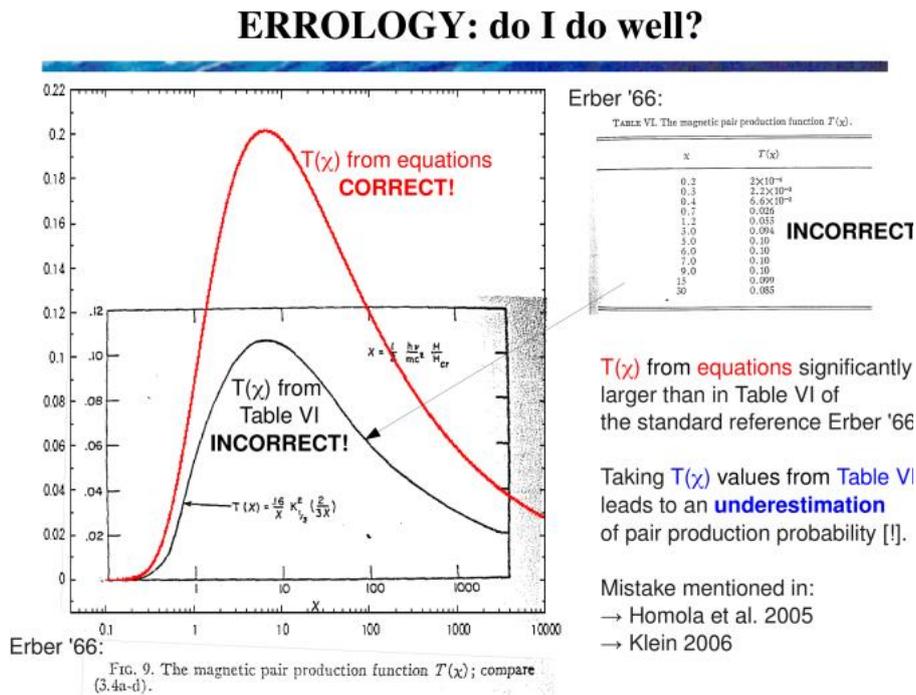

*Fig. 7. An extract from one of talk given by the author mentioning a „factor 2" mistake in the standard reference concerning studies on ultra-high energy photons.*

## 6. Conclusions

Within this article, we argued that *discoverology* might be useful both for scientists and for all science enthusiasts, we also proposed a road map for the proposed new discipline and discussed its major components and milestones, giving practical examples and suggestions. We consider one of these examples particularly valuable: given the availability of the data, the ease of data acquisition and analysis, the most challenging scientific objectives, and the intrinsic need for giant collaboration, we conclude that the CREDO initiative is something more than a scientific project. Already from the above sketch and following the current status and first scientific reports of the CREDO Collaboration, one understands its potential to become a framework for a wide program that could support or trigger new collective, global activities also in steps a), b), d), e), and f) of the *discoverology* road map proposed above. In other words, CREDO has the potential to become a worldwide training ground and standard setter for efforts in the area of *discoverology,* hopefully helping society at large to follow the optimum development path.

*Acknowledgements.*

The author thanks Tamás Csörgő, Teresa Grabińska, Roman Homola, Marlena Jankowska, Johanna F. Jarvis, Michał Jasieński, Łukasz Lamża, Justyna Miszczyk, Tadeusz Pabjan, Magdalena Rzeźnik, and Łukasz Ziobro for critical reading of the manuscript and for their valuable remarks, and colleagues from the Incubator of Scientific Discoveries for inspiring discussions. The author's work was partly funded by the International Visegrad Fund under the grant no. 21720040.